\documentclass[conference, hidelinks, 10pt]{IEEEtran}
\IEEEoverridecommandlockouts

\usepackage{amsmath,amssymb,amsfonts}
\usepackage{algorithmic}
\usepackage{graphicx}
\usepackage{textcomp}
\usepackage[table,xcdraw]{xcolor}
\usepackage{orcidlink}
\usepackage[farskip=0pt]{subfig}
\usepackage{lipsum}
\usepackage[acronym]{glossaries}
\usepackage[detect-all=true,per-mode=repeated-symbol,per-symbol =/]{siunitx}
\usepackage[capitalise]{cleveref}
\usepackage{tablefootnote}
\usepackage{threeparttable}
\usepackage{booktabs}
\usepackage{adjustbox}
\usepackage{multirow}
\usepackage{ragged2e}
\usepackage{microtype}
\usepackage{paralist}
\usepackage{float}
\usepackage{placeins}
\usepackage[backend=biber,style=numeric-comp,sorting=none, maxcitenames=1, mincitenames=1, maxbibnames=1]{biblatex}
\usepackage{geometry}
\usepackage{makecell}
\usepackage{soul}
\usepackage{pifont}

\definecolor{ieee-bright-dblue-100}{rgb}{0.0, 0.3828, 0.6055}
\definecolor{ieee-bright-dblue-80}{rgb}{0.0, 0.4883, 0.6797}
\definecolor{ieee-bright-dblue-60}{rgb}{0.3633, 0.6094, 0.7617}
\definecolor{ieee-bright-dblue-40}{rgb}{0.5898, 0.7383, 0.8398}
\definecolor{ieee-bright-dblue-20}{rgb}{0.8906, 0.8984, 0.9219}
\definecolor{ieee-bright-red-100}{rgb}{0.7266, 0.0469, 0.1836}
\definecolor{ieee-bright-red-80}{rgb}{0.832, 0.3164, 0.3281}
\definecolor{ieee-bright-red-60}{rgb}{0.8906, 0.4922, 0.4805}
\definecolor{ieee-bright-red-40}{rgb}{0.9336, 0.6562, 0.6406}
\definecolor{ieee-bright-red-20}{rgb}{0.9688, 0.8203, 0.8125}
\definecolor{ieee-bright-orange-100}{rgb}{0.9961, 0.6367, 0.0}
\definecolor{ieee-bright-orange-80}{rgb}{0.9844, 0.6953, 0.3125}
\definecolor{ieee-bright-orange-60}{rgb}{0.9883, 0.7695, 0.4844}
\definecolor{ieee-bright-orange-40}{rgb}{0.9922, 0.8359, 0.6562}
\definecolor{ieee-bright-orange-20}{rgb}{0.9961, 0.9219, 0.8164}
\definecolor{ieee-bright-yellow-100}{rgb}{0.9961, 0.8164, 0.0}
\definecolor{ieee-bright-yellow-80}{rgb}{0.9961, 0.8477, 0.2148}
\definecolor{ieee-bright-yellow-60}{rgb}{0.9961, 0.875, 0.4492}
\definecolor{ieee-bright-yellow-40}{rgb}{0.9961, 0.9062, 0.6328}
\definecolor{ieee-bright-yellow-20}{rgb}{0.9961, 0.9531, 0.8125}
\definecolor{ieee-bright-lgreen-100}{rgb}{0.4688, 0.7422, 0.125}
\definecolor{ieee-bright-lgreen-80}{rgb}{0.5742, 0.7852, 0.332}
\definecolor{ieee-bright-lgreen-60}{rgb}{0.6875, 0.8398, 0.5039}
\definecolor{ieee-bright-lgreen-40}{rgb}{0.793, 0.8906, 0.6641}
\definecolor{ieee-bright-lgreen-20}{rgb}{0.8945, 0.9414, 0.8281}
\definecolor{ieee-bright-dgreen-100}{rgb}{0.0, 0.5156, 0.2383}
\definecolor{ieee-bright-dgreen-80}{rgb}{0.1641, 0.6055, 0.3867}
\definecolor{ieee-bright-dgreen-60}{rgb}{0.3906, 0.6953, 0.5234}
\definecolor{ieee-bright-dgreen-40}{rgb}{0.6094, 0.8008, 0.6719}
\definecolor{ieee-bright-dgreen-20}{rgb}{0.8047, 0.8945, 0.8359}
\definecolor{ieee-bright-purple-100}{rgb}{0.5938, 0.1133, 0.5898}
\definecolor{ieee-bright-purple-80}{rgb}{0.6992, 0.3281, 0.668}
\definecolor{ieee-bright-purple-60}{rgb}{0.7812, 0.4961, 0.7461}
\definecolor{ieee-bright-purple-40}{rgb}{0.8555, 0.6602, 0.8281}
\definecolor{ieee-bright-purple-20}{rgb}{0.9219, 0.8281, 0.9023}
\definecolor{ieee-bright-lblue-100}{rgb}{0.0, 0.6094, 0.6484}
\definecolor{ieee-bright-lblue-80}{rgb}{0.0, 0.6797, 0.7188}
\definecolor{ieee-bright-lblue-60}{rgb}{0.2109, 0.75, 0.7812}
\definecolor{ieee-bright-lblue-40}{rgb}{0.5469, 0.8242, 0.8438}
\definecolor{ieee-bright-lblue-20}{rgb}{0.7695, 0.918, 0.9219}
\definecolor{ieee-bright-cyan-100}{rgb}{0.0, 0.707, 0.8828}
\definecolor{ieee-bright-cyan-80}{rgb}{0.0, 0.7227, 0.9453}
\definecolor{ieee-bright-cyan-60}{rgb}{0.2656, 0.7812, 0.957}
\definecolor{ieee-bright-cyan-40}{rgb}{0.5547, 0.8438, 0.9688}
\definecolor{ieee-bright-cyan-20}{rgb}{0.7773, 0.9141, 0.9805}
\definecolor{ieee-bright-white-100}{rgb}{0.9961, 0.9961, 0.9961}
\definecolor{ieee-bright-white-80}{rgb}{0.9961, 0.9961, 0.9961}
\definecolor{ieee-bright-white-60}{rgb}{0.9961, 0.9961, 0.9961}
\definecolor{ieee-bright-white-40}{rgb}{0.9961, 0.9961, 0.9961}
\definecolor{ieee-bright-white-20}{rgb}{0.9961, 0.9961, 0.9961}
\definecolor{ieee-dark-red-100}{rgb}{0.5234, 0.1211, 0.2539}
\definecolor{ieee-dark-red-80}{rgb}{0.6445, 0.2812, 0.3828}
\definecolor{ieee-dark-red-60}{rgb}{0.7422, 0.4727, 0.5234}
\definecolor{ieee-dark-red-40}{rgb}{0.832, 0.6445, 0.6758}
\definecolor{ieee-dark-red-20}{rgb}{0.918, 0.8203, 0.832}
\definecolor{ieee-dark-orange-100}{rgb}{0.9062, 0.4648, 0.1328}
\definecolor{ieee-dark-orange-80}{rgb}{0.9648, 0.5664, 0.3164}
\definecolor{ieee-dark-orange-60}{rgb}{0.9766, 0.6758, 0.4805}
\definecolor{ieee-dark-orange-40}{rgb}{0.9844, 0.7773, 0.6523}
\definecolor{ieee-dark-orange-20}{rgb}{0.9922, 0.8789, 0.8125}
\definecolor{ieee-dark-yellow-100}{rgb}{0.9961, 0.7773, 0.1719}
\definecolor{ieee-dark-yellow-80}{rgb}{0.9961, 0.8086, 0.375}
\definecolor{ieee-dark-yellow-60}{rgb}{0.9961, 0.875, 0.4492}
\definecolor{ieee-dark-yellow-40}{rgb}{0.9961, 0.8984, 0.6875}
\definecolor{ieee-dark-yellow-20}{rgb}{0.9961, 0.9453, 0.8438}
\definecolor{ieee-dark-lgreen-100}{rgb}{0.3945, 0.5508, 0.0938}
\definecolor{ieee-dark-lgreen-80}{rgb}{0.5078, 0.6289, 0.293}
\definecolor{ieee-dark-lgreen-60}{rgb}{0.6367, 0.7188, 0.4688}
\definecolor{ieee-dark-lgreen-40}{rgb}{0.7539, 0.8047, 0.6367}
\definecolor{ieee-dark-lgreen-20}{rgb}{0.875, 0.9023, 0.8125}
\definecolor{ieee-dark-dgreen-100}{rgb}{0.0, 0.3867, 0.2539}
\definecolor{ieee-dark-dgreen-80}{rgb}{0.1836, 0.5, 0.3906}
\definecolor{ieee-dark-dgreen-60}{rgb}{0.3984, 0.6172, 0.5273}
\definecolor{ieee-dark-dgreen-40}{rgb}{0.5938, 0.7422, 0.6758}
\definecolor{ieee-dark-dgreen-20}{rgb}{0.793, 0.8711, 0.8359}
\definecolor{ieee-dark-purple-100}{rgb}{0.4648, 0.1445, 0.5117}
\definecolor{ieee-dark-purple-80}{rgb}{0.5898, 0.3242, 0.6016}
\definecolor{ieee-dark-purple-60}{rgb}{0.6914, 0.4883, 0.6953}
\definecolor{ieee-dark-purple-40}{rgb}{0.7969, 0.6523, 0.793}
\definecolor{ieee-dark-purple-20}{rgb}{0.8945, 0.8203, 0.8945}
\definecolor{ieee-dark-cyan-100}{rgb}{0.0, 0.4492, 0.4648}
\definecolor{ieee-dark-cyan-80}{rgb}{0.0, 0.5469, 0.5664}
\definecolor{ieee-dark-cyan-60}{rgb}{0.3047, 0.6602, 0.668}
\definecolor{ieee-dark-cyan-40}{rgb}{0.5586, 0.7695, 0.7734}
\definecolor{ieee-dark-cyan-20}{rgb}{0.7734, 0.8789, 0.8789}
\definecolor{ieee-dark-dblue-100}{rgb}{0.0, 0.1562, 0.332}
\definecolor{ieee-dark-dblue-80}{rgb}{0.1797, 0.3008, 0.4609}
\definecolor{ieee-dark-dblue-60}{rgb}{0.3828, 0.4609, 0.5859}
\definecolor{ieee-dark-dblue-40}{rgb}{0.5781, 0.6289, 0.7188}
\definecolor{ieee-dark-dblue-20}{rgb}{0.7852, 0.8047, 0.8555}
\definecolor{ieee-dark-grey-100}{rgb}{0.457, 0.4688, 0.4805}
\definecolor{ieee-dark-grey-80}{rgb}{0.5625, 0.5625, 0.5742}
\definecolor{ieee-dark-grey-60}{rgb}{0.6641, 0.6641, 0.6758}
\definecolor{ieee-dark-grey-40}{rgb}{0.7734, 0.7695, 0.7773}
\definecolor{ieee-dark-grey-20}{rgb}{0.8789, 0.8828, 0.8828}
\definecolor{ieee-dark-black-100}{rgb}{0.0, 0.0, 0.0}
\definecolor{ieee-dark-black-80}{rgb}{0.3438, 0.3477, 0.3555}
\definecolor{ieee-dark-black-60}{rgb}{0.5, 0.5078, 0.5195}
\definecolor{ieee-dark-black-40}{rgb}{0.6523, 0.6602, 0.6719}
\definecolor{ieee-dark-black-20}{rgb}{0.8164, 0.8242, 0.8281}

\renewbibmacro{in:}{, }

\DeclareFieldFormat{year}{\addcomma\space#1}
\DeclareFieldFormat[inproceedings]{booktitle}{\textit{#1}\addcomma\space}
    
\renewbibmacro*{issue+date}{%
    \printfield{year}%
    \newunit}

\AtEveryBibitem{%
    \clearfield{title}\clearfield{publisher}\clearfield{location}%
    \clearfield{volume}\clearfield{number}\clearfield{pages}%
    \clearfield{series}\clearfield{note}%
}

\defbibenvironment{bibliography}{}{}
  {\small\noindent\printtext[labelnumberwidth]{%
     \printfield{labelprefix}%
     \printfield{labelnumber}}%
   \addhighpenspace}
\DeclareSIUnit{\x}{\!\ensuremath{\times}}
\DeclareSIUnit\bit{b}
\DeclareSIUnit\flop{FLOP}
\DeclareSIUnit\dash{\text{-}}
\DeclareSIUnit\gateeq{GE}
\newcommand{\vlsitwo}{VLSI 2}
\newcommand{\riscv}{\mbox{RISC-V}}
\newacronym[longplural={systems-on-chip}]{soc}{SoC}{system-on-chip}
\newacronym{asic}{ASIC}{application-specific integrated circuit}
\newacronym{dsa}{DSA}{domain-specific architectures}
\newacronym{ic}{IC}{integrated circuit}
\newacronym{eda}{EDA}{electronic design automation}
\newacronym{fpga}{FPGA}{field-programmable gate array}
\newacronym[longplural={intellectual properties}]{ip}{IP}{intellectual property}
\newacronym{os}{OS}{open-source}
\newacronym{pdk}{PDK}{process design kit}
\newacronym{rtl}{RTL}{register transfer level}
\newacronym{uc}{MCU}{microcontroller}
\newacronym{vlsi}{VLSI}{very-large-scale integration}
\newacronym{isa}{ISA}{instruction set architecture}
\newacronym{sv}{SV}{SystemVerilog}

\newcommand{\x}{$\times$}

\title{\fontsize{16pt}{19pt}\selectfont\bf
Croc: Training the Next Generation Chip Designers on \\ Domain-Specific End-to-End Open Source Silicon
\vspace{-0.3cm}
}

\begin{document}

\author{
    \IEEEauthorblockN{%
    Enrico Zelioli\orcidlink{0009-0005-4899-1047}
    \IEEEauthorrefmark{1}\IEEEauthorrefmark{10}, %
    Philippe Sauter\orcidlink{0009-0001-6504-8086}\IEEEauthorrefmark{1}\IEEEauthorrefmark{10}, %
    Thomas Benz\orcidlink{0000-0002-0326-9676}\IEEEauthorrefmark{1}\IEEEauthorrefmark{2}\IEEEauthorrefmark{10}, %
    Hannah Pochert\IEEEauthorrefmark{1}, \\ %
    Luisa W\"uthrich\IEEEauthorrefmark{1}, %
    Beat Muheim\IEEEauthorrefmark{1}, %
    Frank K. G\"urkaynak\orcidlink{0000-0002-8476-554X}\IEEEauthorrefmark{1}, %
    Luca Benini\orcidlink{0000-0001-8068-3806}\IEEEauthorrefmark{1}\IEEEauthorrefmark{3}%
    }
    \thanks{%
        \IEEEauthorrefmark{10} All authors contributed equally to this research.
    }
    \IEEEauthorblockA{
        \textasteriskcentered~\textit{%
        ETH Zurich}, Switzerland \hspace{0.5cm}
        \textdagger~\textit{%
        lowRISC C.I.C.}, UK \hspace{0.5cm}
        \textdaggerdbl~\textit{%
        University of Bologna}, Italy \\
    }
    \vspace{-1cm}
}

\maketitle

\begin{abstract}
    The demand for domain-specific \glspl{soc} in artificial intelligence, robotics, and automotive systems is increasing the need for engineers with hands-on expertise on \gls{vlsi} design from architecture specification to fabricated silicon.
    Yet, most \gls{vlsi} courses rely on restrictively licensed electronic design automation tools and \glspl{pdk}, as well as closed-source hardware designs.
    We present an end-to-end open-source domain-specific \gls{soc} design and fabrication flow built around Croc, a highly customizable RISC-V platform.
    Built from open-source SystemVerilog intellectual property blocks and integrated with an end-to-end open-source design flow in a 130\,nm open \gls{pdk}, Croc enables tapeout projects supporting multiple domain customization options: instruction-set extensions, accelerator co-processors, and peripherals.
    In our first open-source course experience using Croc, 65 students completed 33 projects, 30 of which produced manufacturable layouts.
    18 designs were selected as tapeout candidates, and five were fabricated.
    A first baseline chip has already been successfully characterized in silicon, demonstrating microcontroller-class functionality and implementation metrics comparable to those of products with similar functional complexity completed with closed-source toolchains and \glspl{pdk}.
\end{abstract}

\glsresetall

\section{Introduction}

The semiconductor industry is accelerating its growth, driven by the increasing demand for specialized \glspl{dsa} for AI, robotics, and automotive applications.
At the same time, recent chip supply chain crises have made clear that the limiting factor is not only fabrication capacity, but also the availability of engineers with practical experience in \gls{ic} design and manufacturing~\cite{2024europeanchipsact}.
This challenge is reflected in recent public initiatives, known as \emph{Chips Acts}~\cite{2024chipsandscienceact, 2024europeanchipsact}, with a strong focus on training, upskilling, and  increasing access to silicon prototyping for the new generations of chip designers.
As demand for silicon continues to increase, this shortage is expected to remain a structural problem in the coming years, leaving a clear educational gap.
To keep up with the recent trends, the semiconductor industry thus needs educational programs that can train a strong and numerous generation of chip designers~\cite{2024edu4chips}.

Many existing \gls{vlsi} courses rely on closed-source \gls{eda} tool licenses, NDA-restricted \glspl{pdk}, requiring a complex security and access-control infrastructure accessible only to a limited number of educational institutions.
As a result, hands-on chip design training remains restricted to a few selected students or universities, while many others only gain theoretical background or \gls{fpga}-based experience~\cite{2024edu4chips}.

Recent advances in \gls{os} hardware platforms, \gls{eda} tools, and open \glspl{pdk}~\cite{sauter2025basilisk} have created a new foundation for chip design training programs that are substantially more accessible and scalable than traditional university offerings.
By eliminating the need for limited tool licenses, NDA-restricted \glspl{pdk}, and closed-source hardware \glspl{ip}, the \gls{os} ecosystem reduces the administrative and accessibility barriers to hands-on RTL-to-GDS education.
This makes it possible to train much larger and more diverse groups of students, including those attending institutions that would otherwise lack the budget, legal framework, and technical infrastructure required for silicon-oriented courses based on confidential closed-source design environments.

In this work, we present the results of the first iteration of the restructured \gls{vlsi} course at ETH Zurich, redesigned in 2025 around a fully open and reproducible teaching framework.
The main goal of the course is to improve students' practical skills and to make hands-on domain-specific chip design education substantially more scalable and transferable beyond a single institution.
The course is built around the \gls{os} Croc \gls{soc} platform~\cite{sauter2025crocendtoendopensourceextensible}. 
Croc's modular architecture is designed to give students fine-grained control over the \gls{soc} configuration and to support domain-specific customizations, namely \gls{isa} extensions, accelerator co-processor integration, and specialized peripherals.
In addition to the \gls{soc} design itself, the Croc platform includes an end-to-end design flow that is beginner-friendly while still allowing extensive customization. The complete teaching material and tool flow are freely accessible under a \emph{Creative Commons} license\footnote{\url{https://vlsi.ethz.ch}}, making the approach substantially more scalable and transferable to other universities and training programs.

\begin{figure*}[t!]
    \begin{minipage}{\linewidth}
        \centering
        \includegraphics[width=\linewidth]{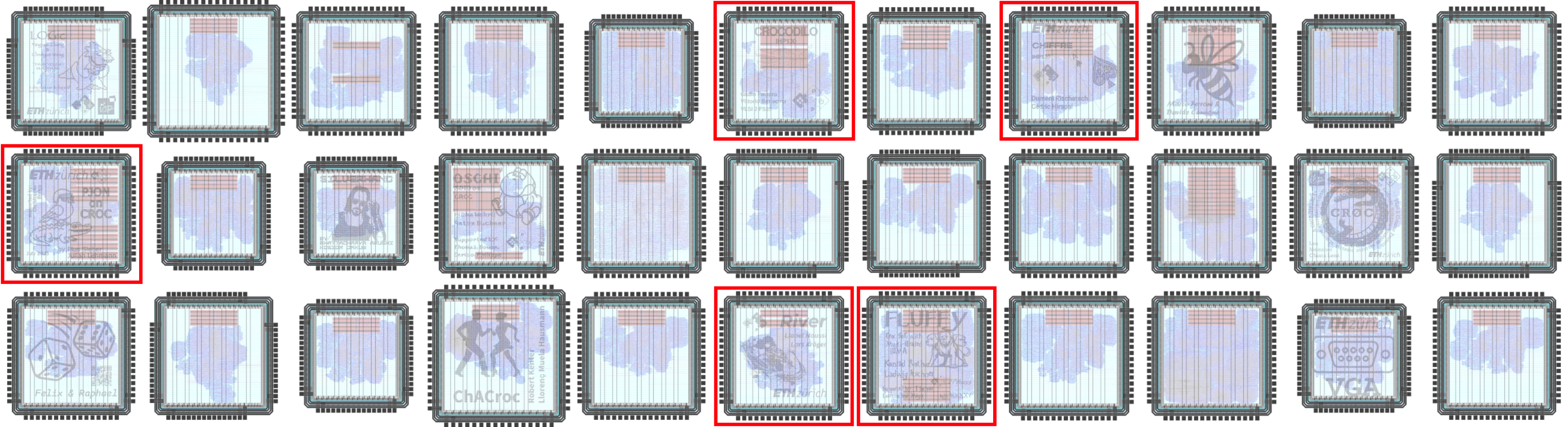}
        \captionof{figure}{\label{fig:die}{GDSII renders of the student layouts produced in the 2025 VLSI course with manufactured ones highlighted.}}
        \vspace{0.25cm}
    \end{minipage}

    \begin{minipage}{1\textwidth}
    \centering
    \begin{minipage}{0.231\linewidth}
        \centering
        \includegraphics[width=\linewidth]{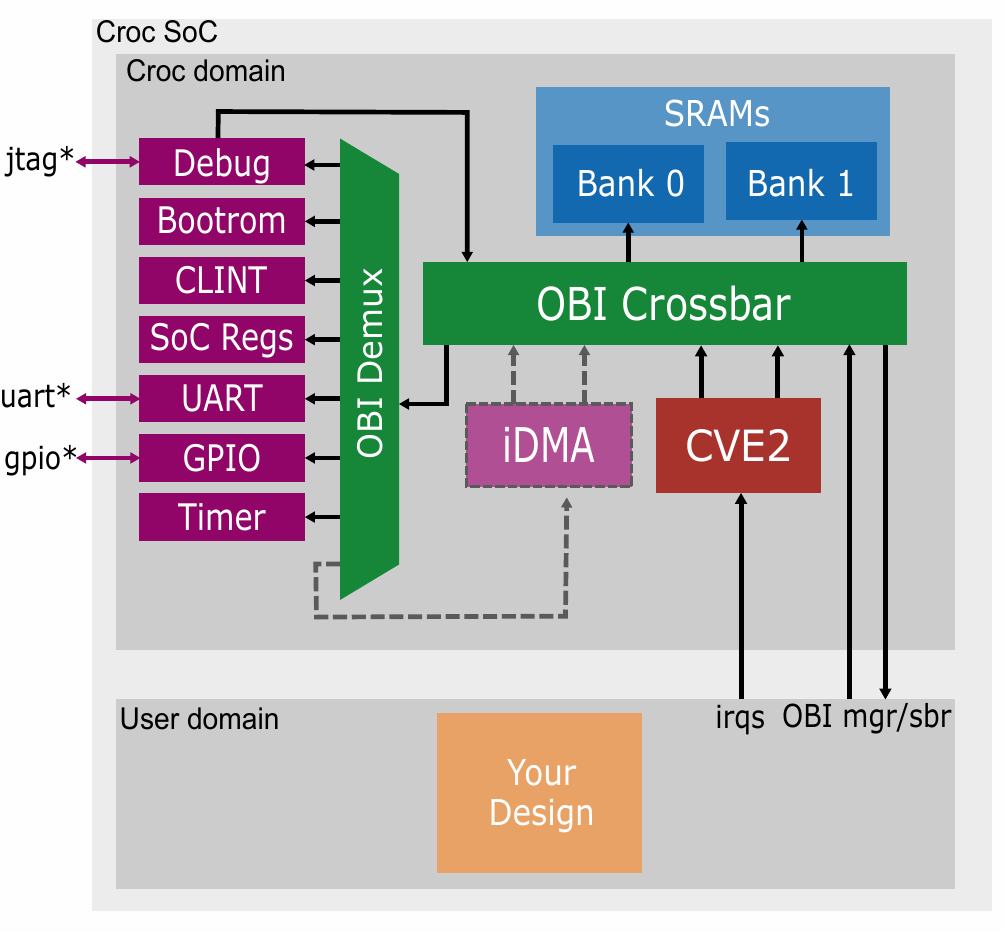}
        \captionof{figure}{Croc's architecture.}
        \label{fig:arch}
    \end{minipage}
    \hfill
    \begin{minipage}{0.237\linewidth}
        \centering
        \resizebox{\columnwidth}{!}{%
    \centering
        \begin{threeparttable}
            \begin{tabular}{@{}l|c|r}
            
                \arrayrulecolor{ieee-dark-black-100}\toprule
                &
                \multicolumn{2}{c}{Num. Groups} \\
                \arrayrulecolor{ieee-dark-black-100}\midrule

                Submitted Design &
                \multicolumn{2}{c}{33} \\

                DRC Clean &
                \multicolumn{2}{c}{30} \\

                Taped Out &
                \multicolumn{2}{c}{5} \\

                \arrayrulecolor{ieee-dark-black-40}\midrule

                Design Addition &
                \multicolumn{2}{c}{ }
                \\

                ~\textbullet~ \gls{dsa} Accelerator &
                \multicolumn{2}{c}{15} \\

                ~\textbullet~ Crypto Accelerator &
                \multicolumn{2}{c}{4} \\

                ~\textbullet~ Core Ext. &
                \multicolumn{2}{c}{7} \\

                ~\textbullet~ Peripheral Ext. &
                \multicolumn{2}{c}{8} \\
                \arrayrulecolor{ieee-dark-black-100}\midrule

                QoR improvements &
                Baseline & Best
                \\

                ~\textbullet~ Max Frequency &
                \SI{80}{\mega\hertz} & \SI{95}{\mega\hertz} \\

                ~\textbullet~ On-Chip Memory &
                \SI{8}{\kilo\byte} & \SI{32}{\kilo\byte} \\

                ~\textbullet~ Placement Density &
                \SI{40}{\percent} & \SI{70}{\percent} \\

                \arrayrulecolor{ieee-dark-black-100}\bottomrule
                
            \end{tabular}

            \begin{tablenotes}[para, flushleft]
            \end{tablenotes}
        \end{threeparttable}
}

        \captionof{table}{Course results.}
        \label{fig:results}
    \end{minipage}
    \hfill
    \begin{minipage}{0.297\linewidth}
        \centering
        \includegraphics[width=\linewidth]{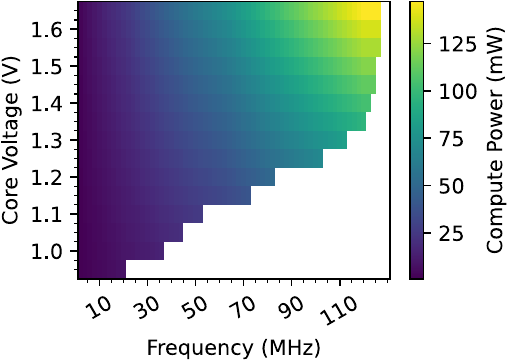}
        \captionof{figure}{Shmoo plot at \SI{25}{\celsius}.}
        \label{fig:shmoo}
    \end{minipage}
    \hfill
    \begin{minipage}{0.215\linewidth}
        \centering
        \includegraphics[width=\linewidth]{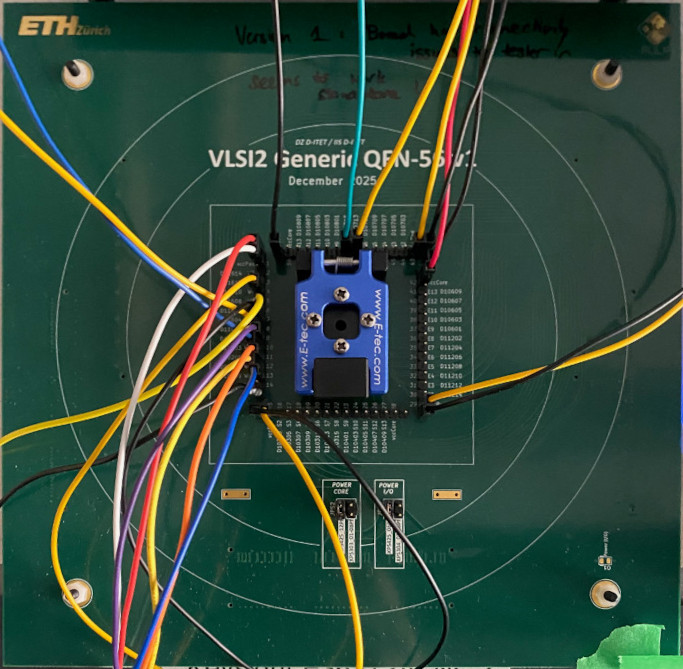}
        \captionof{figure}{MLEM on testbed.}
        \label{fig:testboard}
    \end{minipage}

    \end{minipage}
    \vspace{-0.5cm}
\end{figure*}

\section{The Croc SoC Platform}

Croc features a microcontroller-class {\riscv} \gls{soc} built on \gls{os} silicon-proven SystemVerilog \glspl{ip}.
Croc is designed to be extensible and highly flexible.
Figure \ref{fig:arch} illustrates the top-level architecture of the Croc \gls{soc}.
The design is divided into two main blocks: a \emph{Croc domain}, which contains the basic chip infrastructure, and a \emph{User domain}, which hosts the custom student designs.
Croc is available from a single repository\footnote{\url{https://www.github.com/pulp-platform/croc}} with \gls{rtl}, software setup, and documentation.

The supporting \emph{Croc domain} features an industry-maintained, production-ready \emph{CVE2}~\cite{openhw2022cve2} {\riscv} core, a minimal set of peripherals, and an \emph{OBI}~\cite{silicon2020obi1} crossbar.
The Croc domain primarily facilitates students in implementing, debugging, and verifying their designs.
Students may modify it, for example, by implementing custom \gls{isa} extensions or by extending the interconnect.
Croc features a single-cycle tightly coupled interconnect and two {SRAM} banks, allowing {CVE2} to achieve its ideal performance of one instruction per cycle.
The \emph{User domain} provides an interface for loosely coupled accelerators, co-processors, peripherals, or experimental {\riscv} cores. It exposes the manager and subordinate ports on the main OBI crossbar, along with a set of custom interrupts for communication with the host domain.

Croc's flow targets IHP's \SI{130}{\nano\metre} open \gls{pdk}~\cite{herman2024reflections} with a fully \gls{os} design flow using \emph{Yosys} for synthesis, \emph{OpenRoad} for physical design, and \emph{Verilator} for \gls{rtl} simulation.
The course relies on the \emph{IIC-OSIC-TOOLS}~\cite{iic-jku2024iic-osic-tools} container by JKU Linz to provide the \gls{os} \gls{eda} tools used for exercises.
The Croc platform provides out-of-the-box FPGA emulation for \emph{Digilent}’s \emph{Genesys 2} development board, as well as a \emph{nextpnr}-based flow as an end-to-end \gls{os} alternative~\cite{8735573}.
We collect the silicon results presented in \Cref{sec:results} on a prototype evaluation board running Croc-based \glspl{asic} standalone using DUTCTL~\cite{benz2024dutctl}.
Once the fabricated chips are available, each group with a successful tapeout will receive an evaluation board, currently under development, to support silicon bring-up, measurements and application development on their own chip.

\section{Results}
\label{sec:results}

ETH Zurich has over 35 years of experience in hands-on lectures for \gls{vlsi} design with hundreds of \glspl{asic} designed as part of these courses using commercial \gls{eda}. The first iteration of the revised course demonstrates that first-rate ASIC design courses can be realized effectively using \gls{os} \gls{eda}. During the Spring semester of 2025, 65 graduate and undergraduate students at ETH Zurich completed the course working in teams of two on 33 \gls{asic} projects. Out of these, 30 produced viable layouts, and 18 were selected as tapeout candidates based on their design and implementation improvements.
Available funding allowed five best designs -- including serial communication peripherals, an extended direct memory access engine, and an Elliptic Curve Crypto accelerator -- to be sent for manufacturing in August 2025.
MLEM, a first demonstrator based on the Croc \gls{soc} platform, has been characterized successfully in silicon.
Under typical conditions, our demonstrator runs at \SI{82}{\mega\hertz} and consumes \SI{52.3}{\milli\watt} at a core voltage of \SI{1.2}{\volt} while performing integer computations, which is comparable to performance metrics reported for commercial and advanced research chips implemented in similar technologies~\cite{vivosoc,nxp_lpc111x_2014}.
A power-annotated shmoo plot is shown in \Cref{fig:shmoo}.

\section{Conclusion and Outlook}

We presented the results from the first iteration of our restructured \gls{vlsi} course using an end-to-end \gls{os} reference implementation flow based on the Croc {\riscv} \gls{soc}.
Croc bridges the gap between teaching and industry, allowing students to gain hands-on experience in the design, implementation, and testing of \glspl{ic}.
The measurements show that Croc is suitable to run real microcontroller-grade workloads, and that its basic silicon metrics are comparable to those of commercial and research chips in similar, mature technology nodes.
In 2025, the \gls{vlsi} class started with 65 students at ETH Zurich, producing 30 successful \gls{os} \gls{asic} layouts out of which five \gls{soc} were taped-out using IHP's \SI{130}{\nano\metre} \gls{pdk}.
Together, these results suggest that \gls{os} tools, open \glspl{pdk}, and a carefully designed extensible \gls{soc} template can make hands-on chip-design education substantially more accessible, scalable, and relevant to the industrial practice.

\section{Acknowledgments}
\label{sec:acknowledgments}
\noindent
We thank all students involved in the 2025 {\vlsitwo} class, P. Scheffler and M. Povišer.
We also thank IHP for their generous support.
Silicon manufacturing was sponsored by German BMBF project FMD-QNC (16ME0831).

\printbibliography

\end{document}